\documentclass[a4paper,twoside]{article}

\usepackage{algorithm2e}
\usepackage{apalike}
\usepackage[scaled]{beramono}
\usepackage{booktabs}
\usepackage{calc}
\usepackage{epsfig}
\usepackage[T1]{fontenc}
\usepackage[bottom]{footmisc}
\usepackage{microtype}
\usepackage{multicol}
\usepackage{newtxtext}
\usepackage{newtxmath}
\usepackage{subcaption}
\usepackage{url}
\usepackage{SCITEPRESS}     

\newcommand{\mdx}{mdx\,\uppercase\expandafter{\romannumeral 2\relax}}
\newcommand{\mdxone}{mdx\,\uppercase\expandafter{\romannumeral 1\relax}}

\begin{document}

\title{Performance analysis of \mdx{}: A next-generation cloud platform for cross-disciplinary data science research}

\author{\authorname{Keichi Takahashi\sup{1}\orcidAuthor{0000-0002-1607-5694}, Tomonori Hayami\sup{1}\orcidAuthor{0009-0003-8165-2253}, Yu Mukaizono\sup{1}, Yuki Teramae\sup{1} and Susumu Date\sup{2}\orcidAuthor{0000-0001-7159-289X}}
\affiliation{\sup{1}D3 Center, University of Osaka, Osaka, Japan}
\email{takahashi.d3c@osaka-u.ac.jp}
}

\keywords{Cloud Computing, Data Science, Performance Evaluation, OpenStack}

\abstract{\mdx{} is an Infrastructure-as-a-Service (IaaS) cloud platform designed to accelerate data
science research and foster cross-disciplinary collaborations among universities and research
institutions in Japan. Unlike traditional high-performance computing systems, \mdx{} leverages
OpenStack to provide customizable and isolated computing environments consisting of virtual
machines, virtual networks, and advanced storage. This paper presents a comprehensive performance
evaluation of \mdx{}, including a comparison to Amazon Web Services (AWS). We evaluated the
performance of a 16-vCPU VM from multiple aspects including floating-point computing performance,
memory throughput, network throughput, file system and object storage performance, and real-world
application performance. Compared to an AWS 16-vCPU instance, the results indicated that \mdx{}
outperforms AWS in many aspects and demonstrated that \mdx{} holds significant promise for
high-performance data analytics (HPDA) workloads. We also evaluated the virtualization overhead
using a 224-vCPU VM occupying an entire host. The results suggested that the virtualization overhead
is minimal for compute-intensive benchmarks, while memory-intensive benchmarks experienced larger
overheads. These findings are expected to help users of \mdx{} to obtain high performance for their
data science workloads and offer insights to the designers of future data-centric cloud platforms.}

\onecolumn \maketitle \normalsize \setcounter{footnote}{0} \vfill

\section{\uppercase{Introduction}}\label{sec:introduction}

The rapid advancements in data collection and data analysis capabilities have led to the widespread
adoption of data-driven approaches in scientific research. There is thus an increasing demand within
the academic community for computational infrastructures that facilitate the aggregation, storage,
and analysis of large-scale data. However, the functionalities and performance requirements of such
infrastructures vary significantly across different academic fields and projects, making it
impractical to develop a singular infrastructure tailored to a specific field or project.

To address this situation, research institutions in Japan envisioned a concept of a cloud platform
known as \emph{mdx}. mdx is jointly procured and operated by nine national universities and two
research institutes and provides service to educational and research institutions and private
companies across Japan. The first implementation of the mdx concept, named
\mdxone{}~\cite{Suzumura2022}, was installed at the University of Tokyo and began offering services
to users in September 2021. To enable continued operation during system maintenance or
replacement, and to strengthen disaster resistance and fault tolerance, the second-generation mdx
named \mdx{} was installed at the University of Osaka and started its service in November 2024.

\mdx{} is an Infrastructure-as-a-Service (IaaS) cloud platform for data science research. Unlike
traditional High-Performance Computing (HPC) systems that use batch job schedulers and bare metal
servers, \mdx{} is based on the OpenStack cloud computing platform. This allows users to create
isolated, tailor-made computing environments consisting of virtual machines, virtual storage,
and virtual networks, to meet their diverse compute, storage, and network requirements.
\mdx{} also provides a variety of storage types and access methods to support data ingestion block
storage, parallel file system, and object storage which can be accessed through Lustre, S3 API, and
web interface.

Although at the hardware level \mdx{} resembles traditional supercomputers, its software stack and
use cases are significantly different from supercomputers. In this research, we thus carry out a
comprehensive performance evaluation of \mdx{} to provide current and potential users with excepted
performance characteristics of the system and suggestions to optimize the performance of
High-Performance Data Analytics (HPDA) workloads on \mdx{}. Based on the performance evaluation and
analysis, we also aim to provide feedback on the design and configuration of the \mdx{} system to
its operators, as well as designers of future data-centric cloud platforms.

The rest of this paper is structured as follows. Section~\ref{sec:background} briefly introduces the
overall architecture of the \mdx{} system, and compares it with other academic cloud systems.
Section~\ref{sec:evaluation} carries out a comprehensive performance evaluation and analysis of the
\mdx{} system. Section~\ref{sec:conclusion} concludes this paper and discusses future work.

\begin{figure}
    \centering
    \includegraphics[width=\linewidth]{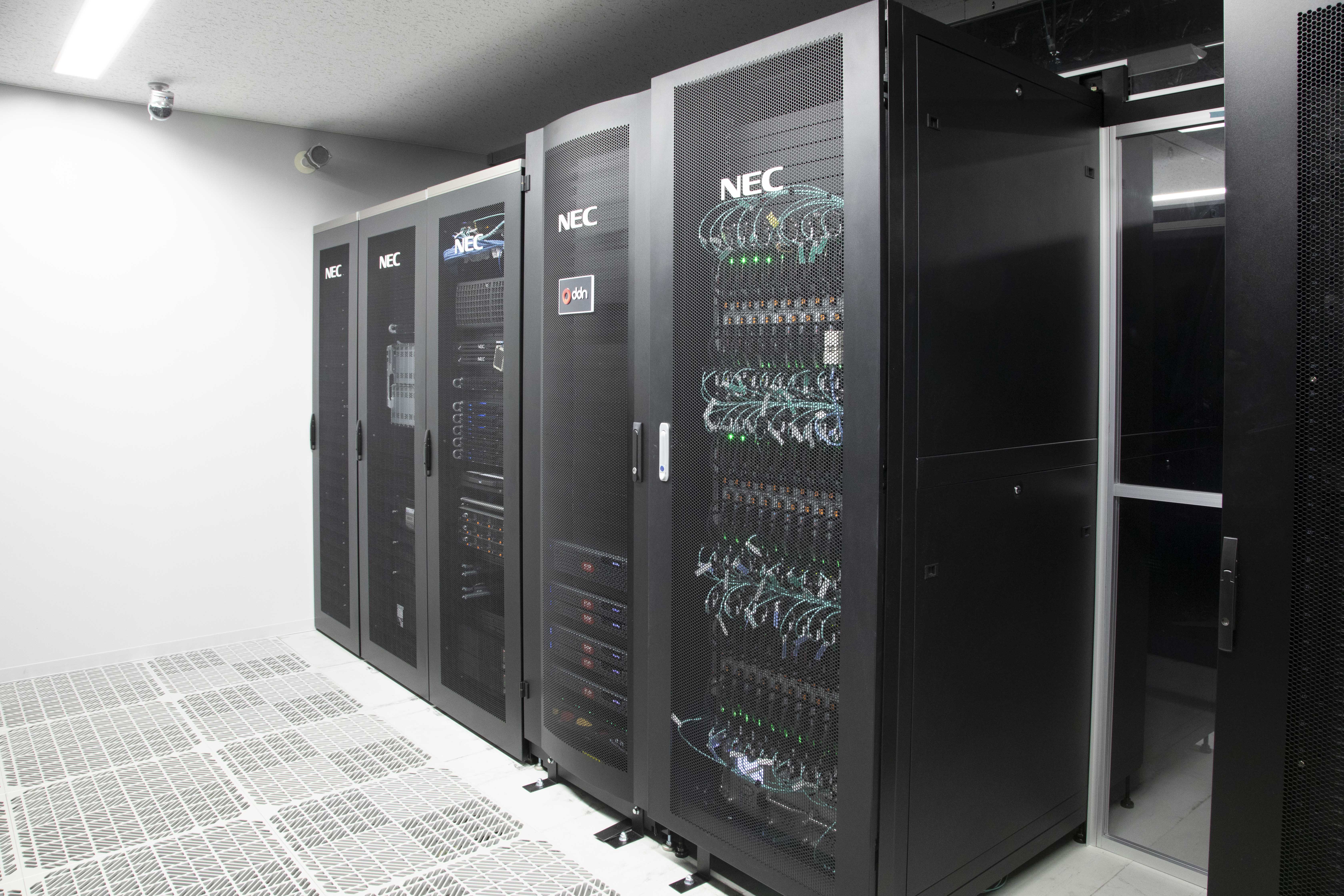}%
    \caption{Server racks installed with servers, storage, and network devices constituting \mdx{}.}\label{fig:rack}
\end{figure}

\section{\uppercase{Background}}%
\label{sec:background}

\subsection{Overview of \mdx{}}

\begin{figure*}
\centering
\includegraphics{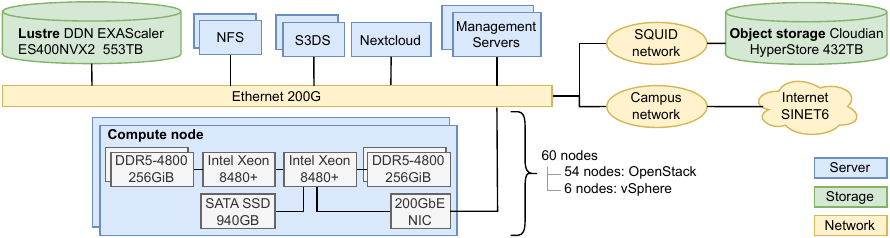}%
\caption{Overall architecture of \mdx{}.}\label{fig:overview}
\end{figure*}

Figure~\ref{fig:overview} shows an overview of the \mdx{} system. \mdx{} comprises 60 compute nodes 
each equipped with two Intel Xeon Platinum 8480+ (Sapphire Rapids) processors and 512\,GiB of
DDR5-4800 SDRAM. Out of the 60 nodes, 54 nodes are managed by the Red Hat OpenStack Platform (RHOSP)
and 6 nodes are managed by VMware vSphere. The vSphere-managed nodes are named
\emph{interoperability nodes}, and allow users to migrate VMs deployed on \mdxone{}, which also
uses vSphere. In this work, we evaluate the RHOSP-managed nodes only.

Two storage systems are provided to compute nodes. One is an all-NVMe Lustre parallel file system
(DDN EXAScaler) with a capacity of 453\,TB, and the other is an S3-compatible object storage
(Cloudian HyperStore) with a capacity of 432\,TB. The Lustre file system can be directly mounted by
multiple VMs used to read and write data from multiple VMs. The Lustre file system is also
used to store VM disk images. Specifically, the Lustre file system is exported as a Network File
System (NFS) volume, which is then accessed by RHOPS's block storage service to create and manage
volumes. In addition, the Lustre file system is also accessible via S3 protocol through the S3 Data
Services (S3DS), which is a Lustre-S3 gateway, and a web interface based on Nextcloud. The
S3-compatible object storage is deployed as a part of an existing supercomputer
(SQUID)~\cite{Date2023}, and connected to \mdx{} via 10\,Gbps Ethernet.

All compute nodes, servers, and Lustre storage are interconnected with a 200\,Gbps Ethernet network.
The overlay network is realized using Generic Network Virtualization Encapsulation
(GENEVE)~\cite{Gross2020}. PCI Passthrough or Single Root I/O Virtualization
(SR-IOV)~\cite{Lockwood2014} are not utilized in \mdx{} because VMs with these technologies cannot
be migrated to other private or public clouds, and one future goal of \mdx{} is to allow seamless
migration of VMs between \mdx{} and public clouds. In terms of external connection, the system is
connected to a Japanese research network (Science Information NETwork, SINET) and the internet
through the University of Osaka's campus network.

\subsection{Related Work}
Jetstream~\cite{Stewart2015} is the first production cloud system within the NSF-funded Extreme
Science and Engineering Discovery Environment (XSEDE)~\cite{Towns2014} ecosystem, designed to
support interactive computing for researchers who do not fit traditional HPC models. Jetstream is an
OpenStack-based cloud system that allows users to provision VMs, and supports authentication and
data movement via Globus. Building on these foundations, Jetstream 2~\cite{Hancock2021} was
introduced as an evolution of the original Jetstream system, featuring heterogeneous hardware
including GPUs, software-defined storage, and container orchestrations.

While \mdx{} and Jetstream share similar goals and both use OpenStack as the cloud computing
platform, an in-depth performance analysis of these systems has not been published to our knowledge.
Except for some performance measurement results obtained as a part of the system acceptance
test~\cite{Stewart2016}, no comprehensive performance evaluation has been conducted, and no
real-world benchmark results or comparisons with public clouds have been published so far.

\section{\uppercase{Performance Analysis}}%
\label{sec:evaluation}

\subsection{Evaluation Method}

As with public clouds, \mdx{} allows users to flexibly choose the number of vCPUs for a VM,
currently ranging from 1 to 224 vCPUs. However, it is infeasible to evaluate all VM
configurations. Thus, in the first half of the evaluation, we focus on a 16-vCPU VM, since the
minimum purchasable vCPU quota is currently 16, and it is expected that many users will launch VMs
of this size. In the second half of the evaluation, we focus on a 224-vCPU VM, since it
occupies a full compute node and thus allows us to directly compare its performance with a bare metal
server that has the same hardware configuration.

\subsection{16-vCPUs VM}

We use the \mdx{} \verb|vc16m32g| instance type, which is equipped with 16 vCPUs and 32\,GiB of
memory, for the evaluation. As a baseline, we use Amazon Web Services (AWS), a widely known public
cloud service. Specifically, we use the \verb|c7i.4xlarge| instance type, which uses CPUs of the
same generation (Sapphire Rapids) as \mdx{} and is equipped with 16 vCPUs and 32\,GiB of memory,
exactly matching the \verb|vc16m32g| instance type. 

\subsubsection{Computing Performance}\label{sec:16vcpu-comp}

We use the Intel-optimized LINPACK Benchmark included in the Intel oneAPI HPC Toolkit 2025.0.1 to
measure the floating-point computing performance and
BabelStream\footnote{\url{https://github.com/UoB-HPC/BabelStream}} 5.0 to measure the memory
throughput. BabelStream is compiled with the Intel oneAPI DPC++/C++ Compiler using the compiler
flags \verb|-O3| \verb|-march=native| \verb|-qopt-zmm-usage=high|
\verb|-qopt-streaming-stores=always|.

Table~\ref{tbl:cm} compares the LINPACK performance and memory throughput between \mdx{} and AWS.
The LINPACK performance of \mdx{} reaches 1.34\,TFLOPS, and is twice that of AWS. This is likely
because \mdx{} does not use vCPU pinning and thus vCPUs are executed on different physical cores.
Contrastingly, AWS pins vCPUs to logical cores to minimize interference with other instances.
Thus, two vCPUs share a single physical core on AWS and results in half the performance of \mdx{}.
The \mdx{} instance achieved 1.7$\times$ higher memory throughput, which could also attribute to a
weaker resource isolation.

\begin{table}
\centering
\caption{Compute and memory performance of a 16-vCPU \mdx{} VM and AWS instance.}\label{tbl:cm}
\begin{tabular}{@{}lrr@{}}
\toprule
\multicolumn{1}{c}{} & \multicolumn{1}{c}{Compute} & \multicolumn{1}{c}{Memory} \\ \midrule
\mdx{}               & 1344 GFLOPS                 & 164 GB/s                   \\
AWS                  & 656 GFLOPS                  & 97 GB/s                    \\ \bottomrule
\end{tabular}
\end{table}

\subsubsection{Network Performance}

To assess the network performance ofa VM, we use iPerf
3.18\footnote{\url{https://github.com/esnet/iperf}} and measure the TCP throughput between VMs
running on either the same compute node or two different compute nodes.
Since a single TCP stream cannot saturate the 200\,Gbps physical
link bandwidth, we generate multiple parallel TCP streams. We also enable the zero-copy option
(\verb|-Z|) in iPerf, which uses the \verb|sendfile()| system call instead of the \verb|write()|
system call to reduce the CPU load and improve the throughput.

\begin{figure}
    \centering
    \includegraphics{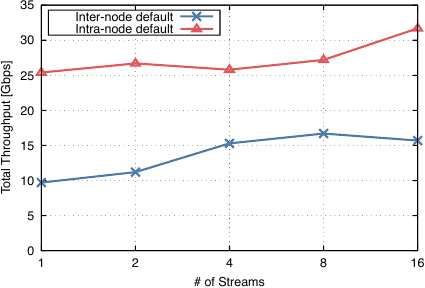}%
    \caption{Total TCP throughput between two VMs.}\label{fig:iperf}
\end{figure}

Figure~\ref{fig:iperf} shows the throughput between two VMs using a varying number of TCP
streams. The single-stream throughput between two VMs running on the same node is 31.6\,Gbps, while
the throughput between VMs running on different nodes is 12.9\,Gbps. As we increase the number of
parallel TCP streams, both the intra- and inter-node throughput does not improve significantly.
This is because by
default, the virtio-net/vhost-net~\cite{Bugnion2017} paravirtualized NIC uses only a single queue to
communicate between the guest and host kernels, and thus all packet transmissions are serialized.
Therefore, the total throughput does not improve with the number of parallel TCP streams.

\begin{figure}
    \centering
    \includegraphics{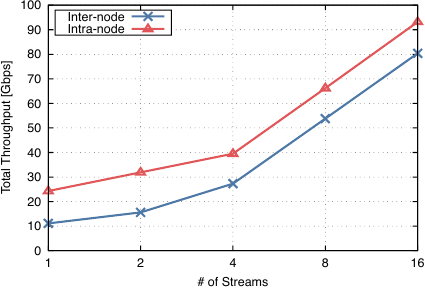}%
    \caption{Total TCP throughput between two VMs with virtio multiqueue enabled.}\label{fig:iperf-mq}
\end{figure}

One optimization to address this problem is to set up multiple queues between the guest and
host kernels. This feature is called \emph{virtio multiqueue}, and can be enabled by setting the
\verb|hw:vif_multiqueue_enabled| flavor extra spec in OpenStack.
Figure~\ref{fig:iperf-mq} shows the throughput with virtio multiqueue enabled. Evidently, the
throughput increases with the number of TCP streams when multiqueue is enabled. With 16 streams, the
inter-node throughput reaches 80.4\,Gbps, and the intra-node throughput reaches 93.2\,Gbps,
demonstrating a significant benefit of virtio multiqueue. Nonetheless, the inter-node throughput is
still lower than the 200\,Gbps link bandwidth of the host. We therefore investigate whether a higher
total throughput can be achieved when multiple VMs simultaneously generate multiple TCP streams.

\begin{figure}
    \centering
    \includegraphics{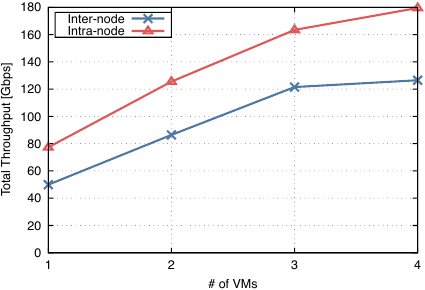}%
    \caption{Total TCP throughput between multiple VM pairs (16 TCP streams per VM).}\label{fig:iperf-parallel}
\end{figure}

Figure~\ref{fig:iperf-parallel} shows the total TCP throughput between multiple pairs of VMs running
either on the same compute node or two different nodes. The number of TCP streams generated by a single VM
is 16. The result shows that the total inter-node throughput increases linearly up to three VMs, and
saturates at 126\,Gbps. We believe the performance gap between the achieved throughput and the
200\,Gbps host NIC bandwidth is due to the various overheads of virtual networking.

In summary, \mdx{} delivers up to 80\,Gbps network throughput to VMs, outperforming most public
cloud instances. This performance, however, is only achievable when virtio multiqueue is enabled.
We are thus recommending the operators of \mdx{} to enable virtio multiqueue by default.

\subsubsection{File System Performance}

Since data science workloads are often I/O-bound~\cite{PhilipChen2014}, the file system performance
of a cloud platform becomes crucial.
Here, we compare the throughput and IOPS of \mdx{} block storage and Lustre storage, and AWS
block storage. We use the Flexible I/O tester (fio)\footnote{\url{https://github.com/axboe/fio}} 3.38 to measure the file access
performance. To saturate the I/O stack, we use the libaio (\verb|--ioengine=libaio|) asynchronous
I/O backend and a sufficiently large (256) number of in-flight I/O requests (\verb|--iodepth|). To
exclude the effect of the page cache, the \verb|O_DIRECT| flag is set (\verb|--direct=1|) to bypass
the page cache. On AWS, we use the General Purpose SSD (gp3) volume type. Since its default I/O
performance is limited (only 3\,KIOPS and 125\,MiB/s throughput), we provision the I/O performance to
its maximum (16\,KIOPS and 1\,GiB/s throughput).

Table~\ref{tbl:io-seq} summarizes the sequential I/O performance. The block storage of \mdx{} offers
4.21\,GB/s in read throughput and 1.75\,GB/s in write throughput. The Lustre storage of \mdx{}
offers 9.82\,GB/s write throughput and 7.74\,GB/s read throughput, surpassing the performance of the
block storage. The OpenStack block storage with the NFS backend works by mounting an NFS volume on the
host, and exposing the image file stored on the NFS volume to the guest via a
virtio-blk~\cite{Russell2008,Bugnion2017} paravirtualized block device. In the case of \mdx{}, the
Lustre-NFS gateway exposes the Lustre volume as an NFS volume. In contrast, Lustre distributes file
accesses to multiple Object Storage Service (OSS) servers and thus provides higher aggregate
performance. The read throughput of Lustre is close to the network throughput measured by iPerf (10.05\,GB/s),
suggesting its is bottlenecked by the guest network performance. The AWS instance exactly delivers
the provisioned 1\,GiB/s performance.

Table~\ref{tbl:io-rnd} summarizes the random I/O performance. Again, AWS exactly delivers the provisioned
16\,KIOPS. The \mdx{} block storage achieves 61\,KIOPS in read and 21\,KIOPS in write access,
exceeding the AWS block storage. The \mdx{} Lustre storage delivers even higher performance of
416\,KIOPS in read and 164\,KIOPS in write access.

To investigate the peak performance of the Lustre storage, we run the
IOR\footnote{\url{https://github.com/hpc/ior}} parallel I/O benchmark on multiple VMs running on different compute nodes,
and measure the sequential read and write performance. 

\begin{figure}
    \centering
    \includegraphics{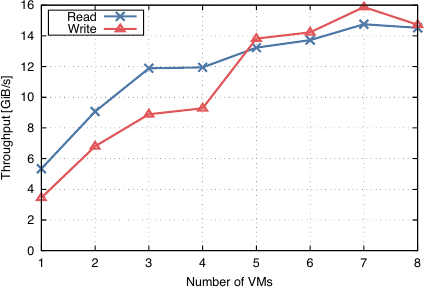}%
    \caption{Total Lustre throughput when accessed from multiple VMs.}\label{fig:lustre}
\end{figure}

Figure~\ref{fig:lustre} shows the total
read and write throughput. The read throughput measured by IOR was 5.33\,GB/s, and the read
throughput was 3.44\,GB/s when the number of VMs was one. The reason the throughput is lower than
the throughput measured by fio is that IOR does not support asynchronous I/O, and thus only one I/O
operation can be in-flight. The total read and write throughput gradually increases with the number
of VMs, and saturates at approximately 15\,GB/s.

In summary, the block storage of \mdx{} outperforms that of AWS, especially in read performance.
Furthermore, the Lustre storage offers considerably higher throughput and IOPS than the block
storage.
It should be noted that this
evaluation was conducted immediately after the launch of the \mdx{} service when system utilization
was still low. Therefore, the I/O performance might become lower due to contention and interference when
the system is highly utilized.

\begin{table}
\centering
\caption{Comparison of sequential I/O performance (1\,MB)}\label{tbl:io-seq}
\begin{tabular}{@{}lrr@{}}
\toprule
\multicolumn{1}{c}{} & \multicolumn{1}{c}{Read} & \multicolumn{1}{c}{Write} \\ \midrule
\mdx{} (Block)       & 4.21 GB/s                & 1.75 GB/s                 \\
\mdx{} (Lustre)      & 9.82 GB/s                & 7.74 GB/s                 \\
AWS (Block)          & 1.05 GB/s                & 1.05 GB/s                  \\ \bottomrule
\end{tabular}
\end{table}

\begin{table}
\centering
\caption{Comparison of random I/O performance (4\,KB)}\label{tbl:io-rnd}
\begin{tabular}{@{}lrr@{}}
\toprule
\multicolumn{1}{c}{} & \multicolumn{1}{c}{Read} & \multicolumn{1}{c}{Write} \\ \midrule
\mdx{} (Block)       & 61 KIOPS                 & 21 KIOPS                  \\
\mdx{} (Lustre)      & 416 KIOPS                & 164 KIOPS                 \\
AWS (Block)          & 16 KIOPS                 & 16 KIOPS                  \\ \bottomrule
\end{tabular}
\end{table}

\subsubsection{Object Storage Performance}

AWS S3-compatible object storages are nowadays widely used as data lakes, and thus many data science
libraries and frameworks support directly loading data from S3-compatible object storage.
To evaluate the performance of the object storages available in \mdx{}, we use
warp\footnote{\url{https://github.com/minio/warp}} 1.0.8, which is a benchmark tool for
S3-compatible storage. We configure warp to either upload or download 2500 objects each of which
is 10\,MiB in size. We vary the number of concurrent operations and measure the total
throughput of GET or PUT operations of the object storage.

\begin{figure}
    \centering
    \includegraphics{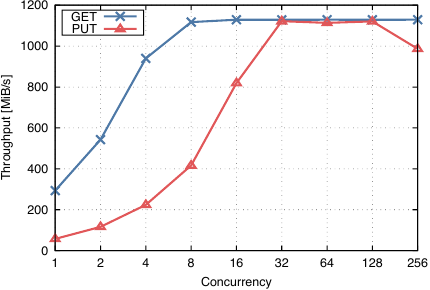}%
    \caption{Cloudian HyperStore throughput.}\label{fig:cloudian}
\end{figure}

Figure~\ref{fig:cloudian} shows the throughput of Cloudian HyperStore. The single-client PUT
throughput is 293\,MiB/s, and the GET throughput is 57\,MiB/s. The throughput increases with the
number of clients, and saturates at approximately 1120\,MiB/s. This is because the HpyerStore object
storage exists on an external supercomputer (SQUID), and the link bandwidth between SQUID and \mdx{}
is 10\,Gbps.

\begin{figure}
    \centering
    \includegraphics{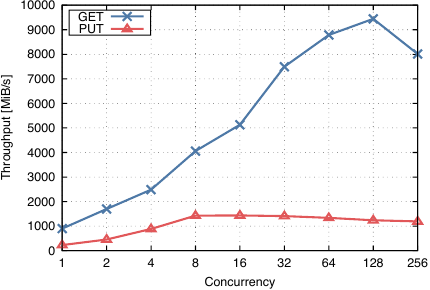}%
    \caption{DDN S3DS throughput.}\label{fig:s3ds}
\end{figure}

Figure~\ref{fig:s3ds} shows the throughput of DDN S3DS. The single-client PUT throughput is
898\,MiB/s and the GET throughput 227\,MiB/s, indicating 3--4$\times$ higher performance than
HyperStore. The GET performance improves considerably with the number of clients and reaches
9444\,MiB/s with 128 clients. This throughput is close to the peak network throughput,
and indicates the guest network performance is the bottleneck. The PUT
throughput saturates at 1435\,MiB/s with 16 clients. This large deviation between the PUT and GET
throughput indicates that S3DS is optimized for GET operations.

\begin{figure}
    \centering
    \includegraphics{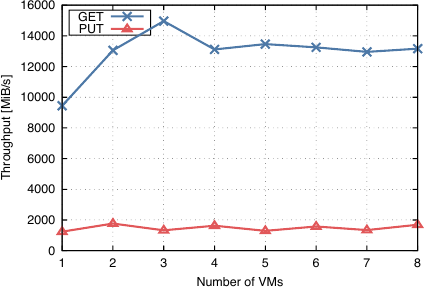}%
    \caption{DDN S3DS throughput when accessed from multiple VMs.}\label{fig:s3ds-multi}
\end{figure}

To evaluate the peak performance of S3DS, we run warp on multiple VMs each running on different
hosts and measure the performance of S3DS when accessed from multiple clients in parallel.
Figure~\ref{fig:s3ds-multi} plots the total throughput with respect to the number of clients (VMs).
The peak GET performance is achieved with 3 clients, and the throughput is 12.75\,GiB/s.
Since this is lower than the Lustre throughput, the S3DS server is the bottleneck. The PUT
performance does not scale with the number of clients, reconfirming that S3DS is optimized for GET
operations.

In summary, Cloudian HyperStore is limited in performance and should be avoided if the workload
requires high S3 access performance. In such cases, the data should be staged to DDN S3DS.


\subsubsection{Data Science Application Performance}

Finally, we assess the real-world performance of \mdx{} in data science workloads.
Since Python its ecosystem remains to be the de facto standard in data
science~\cite{Castro2023,Raschka2020}, we compare the performance of Python-based data science
workloads on \mdx{} and AWS. Specifically, we use the Polars Decision Support (PDS)
benchmarks\footnote{\url{https://github.com/pola-rs/polars-benchmark}}, which is an implementation
of the well-known TPC-H\footnote{\url{https://www.tpc.org/tpch/}} benchmark~\cite{Dreseler2020}
for measuring online analytical processing performance, in different Python libraries
including Pandas, Polars, DuchDB, and Dask. The PDS benchmark allows for running queries on different
dataset sizes. Here, we configure the benchmark to run on a dataset of approximately 10\,GB in size.

\begin{figure}
    \centering
    \subcaptionbox{Pandas}{\includegraphics{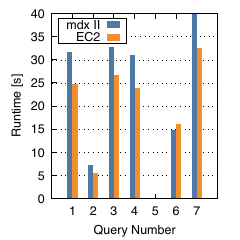}}%
    \subcaptionbox{Polars}{\includegraphics{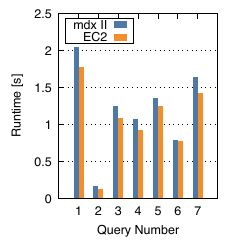}}
    \subcaptionbox{DuckDB}{\includegraphics{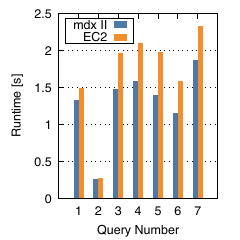}}%
    \subcaptionbox{Dask}{\includegraphics{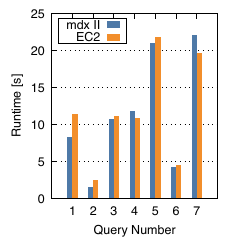}}
    \caption{Polars Decision Support (PDS) benchmarks results.}\label{fig:pds}
\end{figure}

Figure~\ref{fig:pds} compares the runtime for completing each query included in the benchmark suite
on \mdx{} and AWS. The figure shows that there is generally only a small difference in the execution
time of each query between \mdx{} and AWS, but Pandas performs slightly better on AWS and DuckDB
performs better on \mdx{}. However, it should be noted that these performance differences between
\mdx{} and AWS for each library are much smaller than the performance differences between different
libraries. Thus, the choice of the appropriate library for each workload (in this case, Polars or
DuckDB) is critical for maximizing the query performance.

\begin{figure*}
\centering
\includegraphics{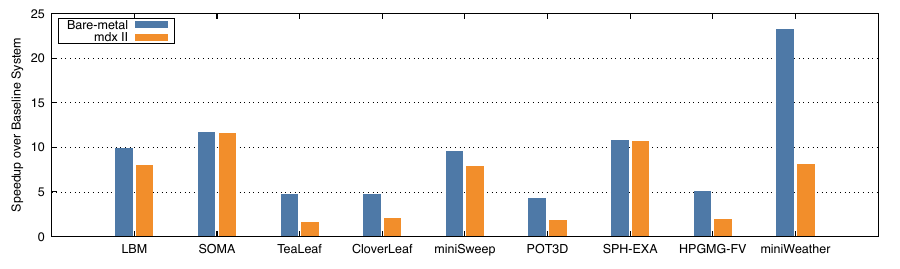}%
\caption{SPEChpc 2021 tiny size results on a bare metal server and mdx II.}\label{fig:spechpc}
\end{figure*}

\subsection{224-vCPU VM}

Since virtualization inherently imposes a performance overhead, we aim to quantify this overhead on
the \mdx{} system in this evaluation. We create a VM occupying a full compute node (\verb|vc224m498g|
instance type) and compare its
performance to a bare metal server with a similar hardware configuration as an \mdx{} compute node.

\subsubsection{Computing Performance}

We first compare the computing performance and memory throughput. Here, the baseline is a
supercomputer named \emph{Laurel} installed at Kyoto University. This system is equipped with two
Intel Xeon Platinum 8480+ CPUs and 512\,GiB of DDR5-4800 SDRAM, matching \mdx{}. Of course, other
components such as storage and interconnect differ from \mdx{}, but we believe their impact in these node-level benchmarks is negligible. The performance is obtained from a previous
work~\cite{Fukazawa2024}.

Table~\ref{tbl:cm-full} summarizes the computing and memory performance. The computing and memory
performance is measured in the same way as the 16 vCPU case described in
Section~\ref{sec:16vcpu-comp}. The result shows that the virtualization overhead imposed to the
floating-point computing performance is minimal. However, the overhead imposed on the memory
throughput is clear. The \mdx{} VM delivers 383\,GB/s memory throughput, which is 22\% lower than
the bare metal server. One reason behind this large performance degradation is the lack of vCPU
pinning in \mdx{}. Because vCPUs are not pinned to host cores, vCPUs can freely move between
the two CPU sockets. This results in a large cross-socket traffic volume in memory-intensive
applications, and degrades the effective memory throughput.

\begin{table}
\centering
\caption{Compute and memory performance of a 224-vCPU \mdx{} VM and a bare metal server}\label{tbl:cm-full}
\begin{tabular}{@{}lrr@{}}
\toprule
\multicolumn{1}{c}{} & \multicolumn{1}{c}{Compute} & \multicolumn{1}{c}{Memory} \\ \midrule
\mdx{}               & 4965 GFLOPS                 & 383 GB/s                   \\
Bare metal           & 4819 GFLOPS                 & 490 GB/s                    \\ \bottomrule
\end{tabular}
\end{table}

\subsubsection{Real-World Application Performance}

To evaluate the impact of virtualization overhead in real-world applications, we use the SPEChpc
2021~\cite{Li2022} benchmark suite.
Various organizations have published the SPEChpc scores of their systems on the SPEChpc official
website\footnote{\url{https://www.spec.org/hpc2021/results/hpc2021tiny.html}}.
In particular, Intel has published measurement results on a server equipped with two Intel Xeon
Platinum 8480+ CPUs and 512\,GiB of DDR5-4800 SDRAM, which matches \mdx{}. 

Figure~\ref{fig:spechpc} compares the performance of the SPEChpc \emph{tiny} suite on the two
systems. The vertical axis shows represents performance of each benchmark, defined as the speedup
over a baseline system (the \emph{Taurus} system at TU Dresden). The plot shows that the
virtualization overhead is relatively small for the LBM, SOMA, miniSweep and SPH-EXA.
On the other hand, the overhead is large for TeaLeaf, CLoverLeaf, POT3D, HPGMG-FV and miniWeather.
TeaLeaf and miniWeather can only achieve one-third of the bare metal performance on \mdx{}. These
benchmarks that experience a large performance degradation on \mdx{} are generally memory-bound, and
thus suffer from the lower memory throughput on \mdx{}.

\section{\uppercase{Conclusions and Future Work}}%
\label{sec:conclusion}

In conclusion, the performance evaluation of the \mdx{} cloud platform demonstrated its superiority over AWS in various metrics, including floating-point computing, memory throughput, and storage I/O performance. This positions \mdx{} as a robust Infrastructure-as-a-Service IaaS solution ideal for HPDA workloads, particularly due to its advanced storage options like Lustre.
While the virtualization overhead is notable in memory-intensive tasks, its minimal impact on compute-intensive benchmarks indicates \mdx{} capability to effectively support diverse HPC and HPDA applications.
Future efforts will focus on analyzing and optimizing the performance of various large-scale real-world data science workloads on \mdx{}, solidifying the role of \mdx{} in advancing data science research and facilitating cross-disciplinary collaborations.

\section*{\uppercase{Acknowledgments}}

This work was partially supported by JST ACT-X Grant Number JPMJAX24M6, as well as JSPS KAKENHI
Grant Numbers JP20K19808 and JP23K16890. The \mdx{} system was used to carry out experiments.

\bibliographystyle{apalike}
{\small
\bibliography{references}}

\end{document}